\documentclass[a4paper,fleqn,usenatbib]{mnras}

\usepackage{newtxtext,newtxmath}

\usepackage[T1]{fontenc}
\usepackage{ae,aecompl}

\usepackage{graphicx}	
\usepackage{amsmath}	
\usepackage{amssymb}	
\usepackage{xcolor}

\newcommand{\beq}{\begin{equation}}
\newcommand{\eeq}{\end{equation}}
\newcommand{\LU}{local $L\sim L_\ast$~sample}
\newcommand{\LS}{$L\sim L_\ast$}

\title[Large Nearby Galaxies]{Formation of the Large Nearby Galaxies}

\author[P. J. E. Peebles]{
P. J. E. Peebles,$^{1}$\thanks{E-mail: pjep@Princeton.edu}
\\
$^{1}$Joseph Henry Laboratories, Princeton University, Princeton, NJ 08544, USA\\
}

\date{Accepted XXX. Received YYY; in original form ZZZ}

\pubyear{2020}

\begin{document}
\label{firstpage}
\pagerange{\pageref{firstpage}--\pageref{lastpage}}
\maketitle

\begin{abstract}
Simulations of galaxy formation tend to place star particles in orbits seriously different from circular in numbers far larger than seem reasonable for the bulges and stellar haloes of  the nearby \LS\ galaxies that can be examined in particularly close detail. I offer an example of how the situation might be improved: a prescription for non-Gaussian initial conditions on the scale of galaxies. 
\end{abstract}




\section{Introduction}

The $\Lambda$CDM cosmology is convincingly established in part because many of its predictions can be computed from first principles by perturbation theory. It is essential that   observations agree with predictions, of course, but equally essential that we can trust the predictions. This is not a value judgement; it is only to say that establishing a persuasive case for the $\Lambda$CDM theory proved to be relatively straightforward.

Galaxy formation cannot be analyzed from first principles.  It is impressive that large-scale numerical simulations based on the $\Lambda$CDM cosmology produce good approximations to real galaxies. And it is inevitable that there are differences between model and observation because galaxy formation is a complex process. The differences are guides to better ways to model the complexity. 

This study is motivated by the possibility that theory and observation disagree in part because some aspect of the physical situation is significantly different from the standard $\Lambda$CDM theory. This cosmology was assembled out of the simplest assumptions I could get away with (Peebles 1982, 1984), and it was not at all surprising to find that some are oversimplifications. An example is the tilt from scale-invariant initial conditions. The length scale issue may be another hint to a better theory (Verde, Treu, \& Riess 2019). The example discussed here is that simulations of galaxy formation tend to produce  a much larger fraction of star particles in far from circular orbits than seems reasonable for a typical  close to pure disc \LS\ galaxy. 

The challenge of reconciling thin galaxies with the hot distributions of orbits found in model galaxies is noted by Kautsch, Grebel, Barazza, \& Gallagher (2006): ``Cosmological models do not predict the formation of disc-dominated, essentially bulgeless galaxies, yet these objects exist.''  Elias, Sales, Creasey, et al. (2018) put it that ``To first order, galaxies with little or no stellar halo are difficult to find in cosmological simulations within $\Lambda$CDM where mergers are a prevalent feature'' but ``{\it How can a galaxy avoid merging and disrupting satellites throughout its entire history?}'' (Italics are in the original).  A discussion of the possible lessons for cosmology seems to be in order. 

Section~\ref{sec:nearby_galaxies} reviews the natures of the stellar haloes and bulge types in the large galaxies that are close enough that they can be examined in best detail. These observations are compared to what might be expected from numerical syntheses of standard ideas about galaxy formation in Section~\ref{sec:theory-observation}. A puzzle that seems to be  related, the nature of the bistability that produced the two great morphological types, spiral and elliptical, is discussed in Section~\ref{sec:bimodality}.  Section~\ref{sec:challenges} presents a summary in the form of five challenges to accepted ideas, along with cautions about the limitations of the evidence. The conclusion offered in Section~\ref{sec:ELS} is that a more promising picture for galaxy formation would be closer to the Eggen, Lynden-Bell \& Sandage (1962) near monolithic collapse. Section~\ref{AdjustingCosmology} offers examples of non-Gaussian initial conditions that change the situation in this direction. Non-Gaussianity is small on scales probed by the cosmic microwave background radiation, but may be significant on the smaller scales of galaxies. For simplicity I use the  warm dark matter initial mass fluctuation power spectrum. Again, it has challenges that may be met by suitably contrived initial conditions. We must be wary of contrived models, but we must be aware of the evidence. 

\section{Stellar Haloes and Hot or Cold Bulges}\label{sec:bulges}

Properties of  classical bulges, pseudobulges, and stellar haloes are well discussed in the literature. This review explains my understanding of issues relevant to this study. 

\subsection{Classical Bulges and Pseudobulges}\label{BulgesPseudobulges}

I take it that a classical bulge of a spiral galaxy is largely supported by a roughly isotropic distribution of stellar velocities, a situation similar to that of an elliptical galaxy. I take it that a pseudobulge is largely supported by a well-ordered flow of stars, perhaps in the manner of the disc stars in a spiral galaxy. The more cautious Kormendy \& Kennicutt (2004) statement is that pseudobulges may ``have one or more characteristics of discs [such as] large ratios of ordered to random velocities.'' My working assumption is that this large ratio is the defining feature. Cold streaming flow in a pseudobulge would allow formation of  the observed spiral arms, rings, and bars, just as for the cold flow of stars in a disc. A classical bulge may be significantly flattened by angular momentum. The same is true of an elliptical galaxy of similar luminosity (Davies, Efstathiou, Fall, \& Schechter 1983; de Zeeuw \& Franx 1991). The hot distribution of orbits in an elliptical tends to discourage pattern formation, but ellipticals can exhibit shells presumed to be sheets in phase phase produced by dry mergers. The hot orbits in classical bulges seem to make them featureless, but I am not aware of a search for shells.

The categories, hot and cold, tell us something about how bulges formed. Stars that formed in subhaloes before merging with the protogalaxy are not likely to have joined a cold distribution of orbits in a pseudobulge. Such pseudobulges more likely formed from gas and plasma that settled to support by streaming flow before being incorporated in stars. Classical bulges could have grown out of stars that formed in subclusters before merging, if in concentrations compact enough to have resisted tidal disruption until joining the bulge. Or diffuse matter may have tumbled toward the center of the growing galaxy and collapsed to stars before it could have settling to organized streaming motion. Or classical bulge stars may have formed in a disc that contracted to a bar that was so violently unstable that gravity rearranged the stars into the cuspy radial distribution characteristic of a classical bulge. I take the point to be that classical bulges likely formed in conditions far from dynamical equilibrium, while pseudobluges likely formed in conditions close to organized flow in near dynamical equilibrium.

\subsection{The Circularity Parameter}\label{sec:circularities}

Model galaxy discs and bulges often are characterized by the distributions of the circularity parameter $\epsilon$ of a star particle orbit,
\beq
\epsilon = {J_z/J_c}.\label{eq:circularity}
\eeq
The component of the angular momentum of the star particle normal to the disc is $J_z$, and $J_c$ is the angular momentum of a particle in a circular orbit in the plane of the disc with the same energy as the star particle. 

Abadi, Navarro, Steinmetz, \& Eke (2003) introduced the elegant decomposition of the frequency distribution of $\epsilon$ into a spheroid component centered near $\epsilon=0$ that could include a classical bulge and stellar halo; a thick disc that would have values of the circularity parameter closer to $\epsilon=1$; and a thin disc with circularity parameters  quite close to $\epsilon=1$. Other early applications of this statistic include Governato, Willman, Mayer, et al. (2007) and, in a variant of this statistic, Scannapieco, Tissera, White, \& Springel (2008). The distribution of circularity parameters in the Guedes, Mayer, Carollo, \& Madau (2013) Eris simulation (in the middle panel in their fig.~5) has a prominent peak at $\epsilon = 1$ from star particles that are in close to circular orbits, as in a disc, and a local maximum at $\epsilon=0$, as from stars in a slowly rotating classical bulge or stellar halo. Most Eris star particles in the range $\epsilon\la 0.8$ are within 2~kpc of the center. They are moving in the direction of rotation of the disc but with considerable departures from circular motion. I take it that they are good candidates for the stars in a classical bulge.

\begin{figure}
\begin{center}
\includegraphics[angle=0,width=2.5in]{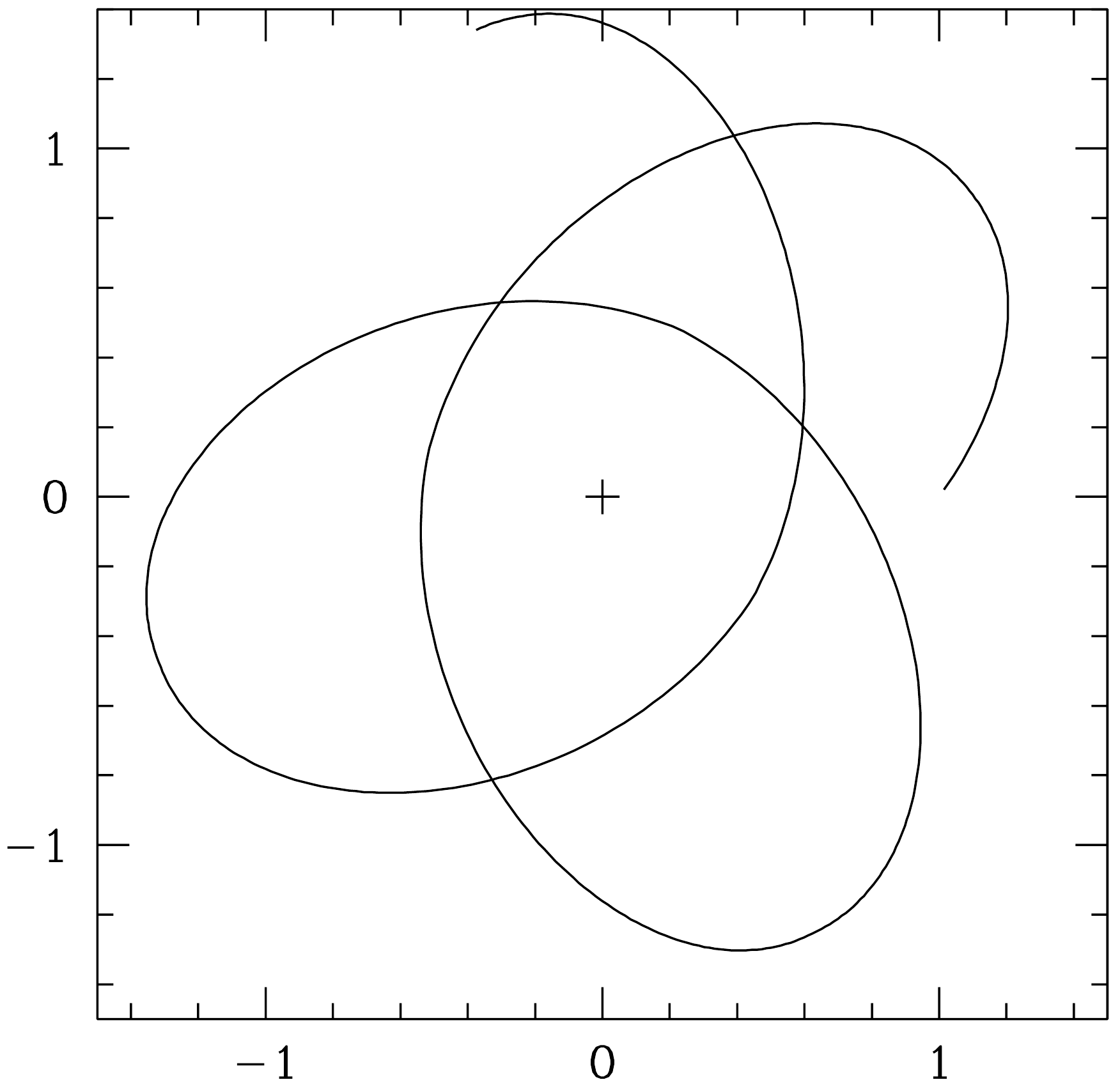} 
\caption{Orbit of a particle with circularity parameter $\epsilon = 0.8$ in the plane of an axisymmetric disc with a flat rotation curve. The circular orbit with the same energy and angular momentum has unit radius.}\label{fig:orbit}
\end{center}
\end{figure}

It is helpful to have illustrations of the relation between values of $\epsilon$ and the natures of the orbits. Consider a model galaxy with an axially symmetric mass distribution and a flat rotation curve with circular speed $v_c$. The galactocentric radius $r$ and angular position $\theta$ of a particle moving in the plane of symmetry  at circularity parameter $\epsilon$ satisfy 
\beq
(dr/dt)^2 = v_c^2\left(1 - 2\log r/R - (\epsilon R/r)^2\right),\ 
d\theta/dt=\epsilon R v_c/r^2\label{eq:orbiteq}
\eeq
A circular orbit with the same energy has radius $R$.

If $\epsilon$ is close to unity then in lowest nonzero order in perturbation theory the maximum departures from a circular orbit, where $dr/dt$ vanishes, satisfy
\beq
r_x = R(1\pm\delta_x), \quad \delta_x^2 = (1-\epsilon^2)/2. \label{eq:deltax}
\eeq
At $\epsilon=0.8$ the extrema of the galactocentric distances are $r=R(1\pm 0.4)$, in this approximation. 

The extrema of the radial velocity are 
\beq
v_{r,\rm max}/v_c=\pm\sqrt{-2\log\epsilon} \hbox{ at } r=\epsilon R. \label{eq:rdot}
\eeq
Another simple measure is the radial velocity at $r=R$:
\beq
dr/dt = \pm v_c\sqrt{1-\epsilon^2} \hbox{ at } r= R. \label{eq:rdot2}
\eeq
At $\epsilon = 0.8$ both velocities are on the order of 60\% of $v_c$. 
 
A numerical solution to equation~(\ref{eq:orbiteq}) is shown in Fig.~\ref{fig:orbit}. The computed maximum and minimum radius and the rms radial velocity at this value of the circularity parameter, $\epsilon = 0.8$, are
\beq
r_{\rm max}=1.40R,\quad r_{\rm min}=0.53R,\quad 
\langle (dr/dt)^2\rangle^{1/2} = 0.43 v_c. \label{eq:orbitexample}
\eeq
At $\epsilon = 0.9$ the numbers are
\beq
r_{\rm max}=1.30R,\quad r_{\rm min}=0.67R,\quad 
\langle (dr/dt)^2\rangle^{1/2} = 0.31 v_c. \label{eq:orbitexample2}
\eeq
These numerical results are reasonably consistent with the approximation  in equation~(\ref{eq:deltax}) and the characteristic velocities in equations~(\ref{eq:rdot}) and (\ref{eq:rdot2}). 

Under the assumption of a spherically symmetric mass distribution, which might be a reasonable approximation where the mass is dominated by the dark halo, we can let the orbit in Fig.~\ref{fig:orbit} at $\epsilon = 0.8$ in the plane of the disk be tilted from the (massless) disc by angle $\theta = 25^\circ$. Then with equation~(\ref{eq:orbitexample}) we see that the orbit rises normal to the disc by the maximum distance $h=r_{\rm max}\sin\theta = 0.6R$. The eccentricity parameter in this tilted example is $\epsilon = 0.8\cos(\theta)=0.7$. A common choice of the disc particles in a model galaxy is $\epsilon \ge 0.7$. 

The point of this simple model is that a distribution of orbits with $\epsilon\sim 0.7$ to 0.9 includes fractional departures $\sim 30$ to 40 per cent from circular motion in radius and radial velocity, and possibly rising out of the disc by like amounts. This may be reasonable for the motions of stars in a classical bulge or in a bar in the disk. But it is not suggested by observations of close to pure disks in nearby \LS\ galaxies, and it is not suggested by the considerations of their pseudobulges in Section~\ref{BulgesPseudobulges}. This discussion continues in Section~\ref{sec:epsilon_distributions}.

\subsection{Other Measures}\label{sec:bulge_measures}

Bulge types may be defined by the fit of the surface brightness $i$ normal to the disc as a function of galactocentric radius $r$ to the sum of an exponential to represent the disc and a S\'ersic function to represent the bulge component,
\beq
\log i = A -Br  - C r^{1/n}. \label{eq:Sersic}
\eeq 
The constants A, B, C, and n are fitting parameters. De Vaucouleurs (1948) introduced the last term with $n=4$ to describe the surface brightness run in an early-type galaxy; S\'ersic (1963) proposed the generalization to a free value of $n$; Freeman (1970) pioneered application of the sum of de Vaucouleurs' form and the exponential to measured surface brightness runs in spirals, and Andredakis, Peletier, \& Balcells (1995) used the generalization to the free parameter $n$ in equation~(\ref{eq:Sersic}). A fit to the measurements with S\'ersic index $n>2$, meaning a cuspy central concentration of starlight, is taken to indicate a classical bulge. A pseudobulge defined by this fit has S\'ersic index $n<2$, meaning a less cuspy central peak, maybe approaching the exponential $n=1$ characteristic of the radial distribution of disc stars. 

Gadotti (2009) introduced another bulge classification. Classical bulges are taken to be those with bulge half-light radius $r_e$ and average surface brightness $\langle\mu _e\rangle$  consistent with the correlation  observed for elliptical galaxies, and pseudobulges are taken to be the lower surface brightness outliers.  Gadotti's measure is a helpful indicator that elliptical galaxies are more similar to classical bulges than pseudobulges. The S\'ersic index $n$ is a helpful indicator of whether the bulge star distribution bears a natural relation to the disc, with S\'ersic index closer to $n=1$, or more naturally to an elliptical-like bulge with index closer to $n=4$. Gadotti (2009) applied this criterion to SDSS images. Gao, Ho, Barth, \& Li (2020) used it in  analysis of the Carnegie-Irvine Galaxy Survey (CGS).

As remarked above, another kind of pseudobulge may be indicated by the presence of a bar that grew out of the instability of a cold circular flow of disc stars. 

I do not know how reliably the nature of bulge support, whether hot or cold, can be determined by these or other indicators applied to real galaxies. But close observations of the nearby \LS\ galaxies seem to be a promising place to start. 

\section{Bulges of Nearby Large Galaxies}\label{sec:nearby_galaxies}

Table~1 shows measurements of the ratios B/T of bulge to total luminosities of 34 galaxies at distances less than 10~Mpc and luminosities \LS. It includes the 27 galaxies with K-band luminosities $L_{\rm K} >  10^{10}$ and distances $D\leq 10$~Mpc  in the Local Universe catalog maintained by Brent Tully et al.\footnote{Available at the Extragalactic Distance Database, http://edd.ifa.hawaii.edu as the catalog `Local Universe (LU)'.} The other 7 galaxies with absolute magnitudes $M_B<-19.6$ are drawn from the list in Fisher \& Drory (2011). The  different wavelengths mean different bounds on stellar masses. The  most extreme case, NGC\,2787, has K-band luminosity above the cut and B-band luminosity well below the optical cut. The situation may be related to the impression of considerable dust across the face of this galaxy. The other galaxies with listed values of $L_K$ and $M_B$ would be included by either cut, or reasonably close to it.

The last column in Table~1 is the bulge type, C or P for classical or pseudobulge, and the source for the bulge type and measured B/T:  K for Kormendy, Drory, Bender, \& Cornell (2010) and F for Fisher \& Drory (2011). Where both papers give a measurement of B/T I use the Kormendy et al. value. The two are not very different. The 19 measurements from Kormendy et al. were meant to be a reasonably complete sample to 8~Mpc distance. The Fisher \& Drory measurements are largely at distances 8 to 10~Mpc, the few nearer than 8~kpc a result of the use of different distance measurements.  

 \begin{table}
 \centering
 \caption{The Nearby Sample of \LS\ Galaxies}
 \label{tab:nearbysample}
\begin{tabular}{lccc}
\hline
galaxy  & D, Mpc  & B/T & type  \\
\hline
Milky Way & 0.01 & $0.19 \pm 0.02$ & P K  \\
M31 & 0.8 & $ 0.32 \pm 0.02 $ & C K \\
NGC4945 & 3.4 & $0.073 \pm 0.012$ & P K   \\
Maffei1 & 3.4 & $1$ & C K  \\
IC342 & 3.4 & $0.030 \pm 0.001$ & P K \\
Maffei2 & 3.5 & $0.16 \pm 0.04$ & P K \\
CenA  & 3.6 & $1$ & C K \\
M81 & 3.6 & $0.34 \pm 0.02$ & C K  \\
NGC253 & 3.7 & $0.15$ & P K \\
Circinus  & 4.2 & $0.30 \pm 0.03$ & P K \\
M64 & 4.4 & $0.20 $ & CP K \\
M94 &  4.4& $0.36 \pm 0.01$ & P K \\
M83 & 4.9 & $0.074 \pm 0.016$ & P K \\
NGC6946 & 6.2 & $0.024 \pm 0.003$ & P K\\
NGC3621 & 6.6 & $0.01$ & P F  \\
M101 & 7.0 & $0.027 \pm 0.008$ & P K \\
M96 & 7.2 & $0.26$ &  P F \\
NGC2787& 7.5 & $0.39$ & CP K\\
M106 & 7.6 & $0.12 \pm 0.02$ & C K  \\
NGC2683 & 7.7 & $0.05 \pm 0.01$ & C K \\
M63 & 7.9 & $0.19$ & P F  \\ 
M66 & 8.3 & $0.10$ & P F \\
M51 & 8.4 & $0.095 \pm 0.015$ & P K \\
NCG2903 & 8.5 & $0.10$ & P F \\
M74 & 9. & $0.08$ & P F  \\
NGC4096 & 9.2 & $0.08$ & P F \\
NGC6744 & 9.2 & $0.15$ &  C F\\
NGC925 & 9.4 & $0.07$ & P F \\
M108 & 9.6 & $0.21$ & P F \\
Sombrero & 9.8 & $0.51$ &  C F \\
NGC3344  & 9.8 & 0.08 & C F \\
M105 & 10. & $1.$ & C F \\
M95 & 10. & $0.16 $ & P F \\
M65 & 10. & 0.16 & P F \\
\hline
\end{tabular}
\end{table}

Kormendy et al. give luminosities of both a classical bulge and a pseudobulge in M94 and NGC\,2787; the sums of the luminosities are entered in Table~1. I do not include three galaxies with Kormendy et al. B/T measurements: M82, because the definition of its stellar bulge seems likely to be awkward;  M51b, which is the spread of stars at the edge of the otherwise elegant M51a spiral galaxy; and NGC\,4490, which looks like a galaxy in the process of merging or falling apart. Six of the LU galaxies that pass the cuts $D<10$~Mpc and $L_K>10^{10}$ do not have bulge luminosities listed in either source. Three of them, NGC\,2640, NGC\,1023, and NGC\,2784,  look like early types. Three, NGC\,4517, NGC\,4631, and NGC\,891, look like normal close to edge-on discs in which dust may or may not obscure significant bulge luminosities. But these six galaxies do not seem to be seriously atypical of what I take to be the \LU. 

There are no super-luminous galaxies $L\sim 10 L_\ast$ in this sample, but they are rare. Since $L\sim L_\ast$ galaxies contribute most of the global starlight it is customary and reasonable to consider them the dominant galaxy class. I have aimed to make Table~1 a reasonably close to fair sample of these dominant galaxies outside rich clusters. It may be well to bear in mind that I cannot offer much evidence that this is really so. But we do know that the bulges of the galaxies in Table~1 have been characterized by seasoned observers. 

\begin{figure}
\begin{center}
\includegraphics[angle=0,width=2.5in]{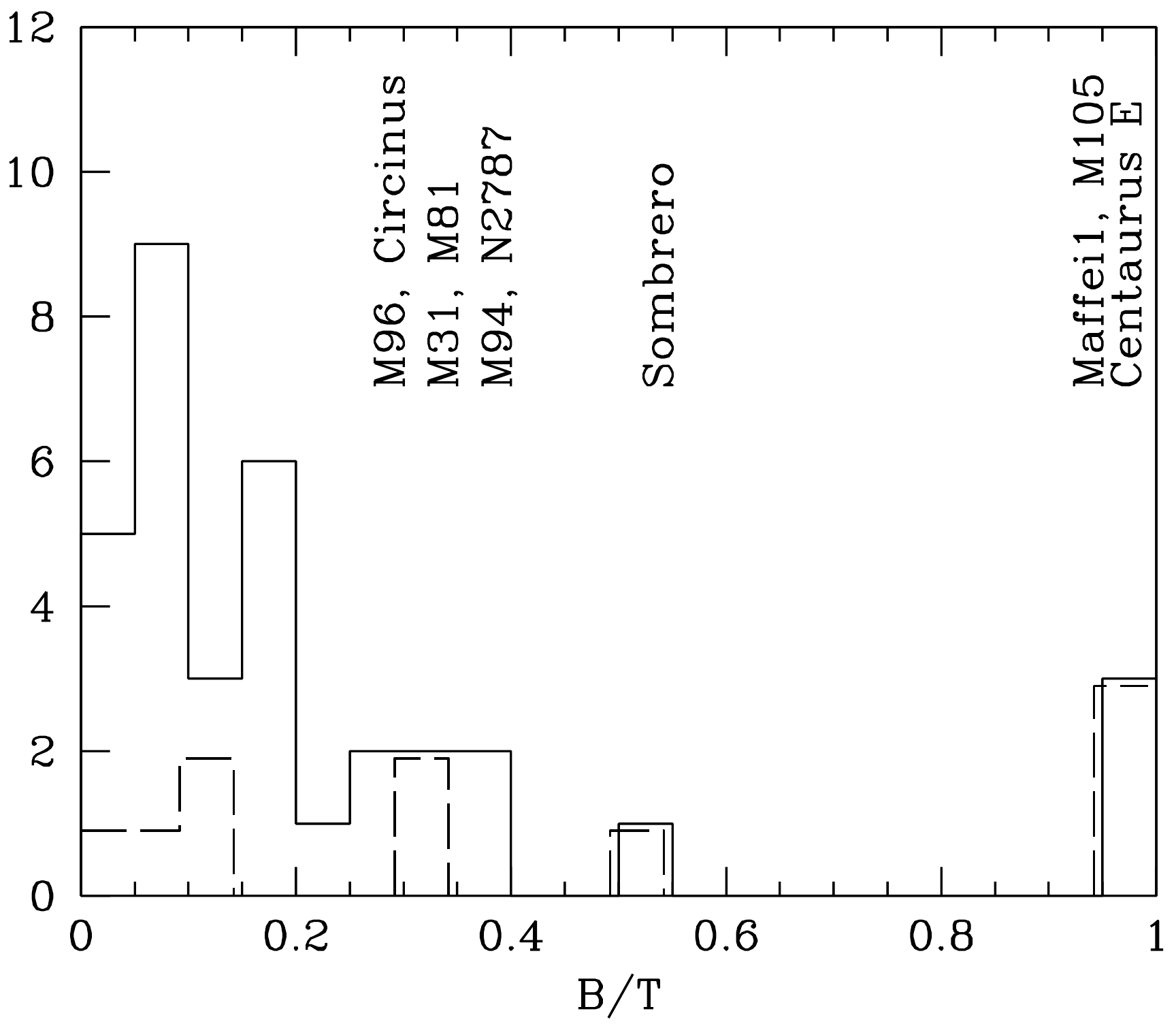}
\caption{Distribution of ratios of bulge to total luminosities in the \LU. The dashed lines shows the subset of classical bulges.}\label{fig:B/Tdata}
\end{center}
\end{figure}

Fig.~\ref{fig:B/Tdata} shows the distribution of ratios B/T of bulge to total luminosities in the \LU. The galaxies with larger values of B/T are named. The three with B/T close to unity are ellipticals. Half the luminosity of the Sombrero Galaxy M\,104 is from a disc, the other half from a stellar bulge or halo. The Andromeda Nebula M\,31 and the spiral M\,81 have relatively prominent classical bulges. But to my mind the striking observation is the considerable fraction of nearby $L\sim L_\ast$ galaxies that are close to pure discs.  An example is M\,101. Kormendy et al. (2010) showed, and  Peebles (2014, fig. 3) used their results to illustrate, the spiral pattern that runs quite close to the $\sim 10^6M_\odot$ nuclear star cluster. This pattern suggests the stellar velocity dispersion relative to the mean in the inner parts of this galaxy is small enough to allow the formation of features, down to the nuclear star cluster that I suppose is supported by near isotropic motions. We might say this galaxy has a classical  bulge with luminosity four orders of magnitude less than the disc. Fisher \& Drory (2008) present HST images of other examples of the fascinating phenomenon of compact nuclear star clusters. 

We can compare three measured distributions of the luminosity fraction B/T in galaxies with pseudobulges. Figure~\ref{fig:B/Tdata} shows the distribution in the local sample. Most are pseudbulges; the dashed lines show the small subset of classical bulges. The other two samples use Gadotti's (2009) criterion: pseudobulges are defined by a departure downward in surface brightness from the relation between effective radius and surface brightness for elliptical galaxies. Gadotti's distribution (lower left-hand panel in fig.~8) is based on SDSS images. The Gao, et al. (2020 fig. 2d) distribution is based on their CGS sample. Both peak at B/T $\la 0.1$, with modest tails to B/T $\sim 0.5$, consistent with the local sample. A  larger fraction of the CGS galaxies are found to have classical bulges than in the local sample, but the criteria differ (Sec.~\ref{sec:bulges}).

Ogle, Lanz, Appleton, et al. (2019) examined the 1500 most luminous galaxies, $L\sim 10L_\ast$, detected in the Sloan Digital Sky Survey catalog of some one million galaxies (Simard, Mendel, Patton, et al., 2011). They identify 84  giant spirals, about 6\% of the most luminous galaxies.  Most of the rest are elliptical or cD galaxies. The median bulge luminosity fraction in the giant spirals, from the Simard et al. fit of the surface brightness run to equation~(\ref{eq:Sersic}) with $n=4$, for the 81 giant spirals with measured bulges, is B/T$= 0.17$ (Patrick Ogle, private communication). This is comparable to the the median for the 31 local \LS\ spiral galaxies, B/T$= 0.15$. 

Ogle's assessment is that better data are needed to determine whether giant spirals have pseudobulges. But there is a good case that the median ratio of bulge to total luminosity of field spiral galaxies with optical-near infrared luminosities $L_\ast\la L\la 10L_\ast$ is B/T$= 0.16\pm 0.03$.

\section{Comparisons of Models and Observations}\label{sec:theory-observation}

It is understood that numerical simulations of galaxy formation are a work in progress aimed at exploring how best to model the complexities of stellar formation and feedback while striving for ever better mass and position resolution. Discrepancies between theory and observation may only indicate the need for more work. But the thought pursued here is that discrepancies may suggest clues to a better cosmology to serve as the basis for simulations. 

\begin{figure}
\begin{center}
\includegraphics[angle=0,width=2.5in]{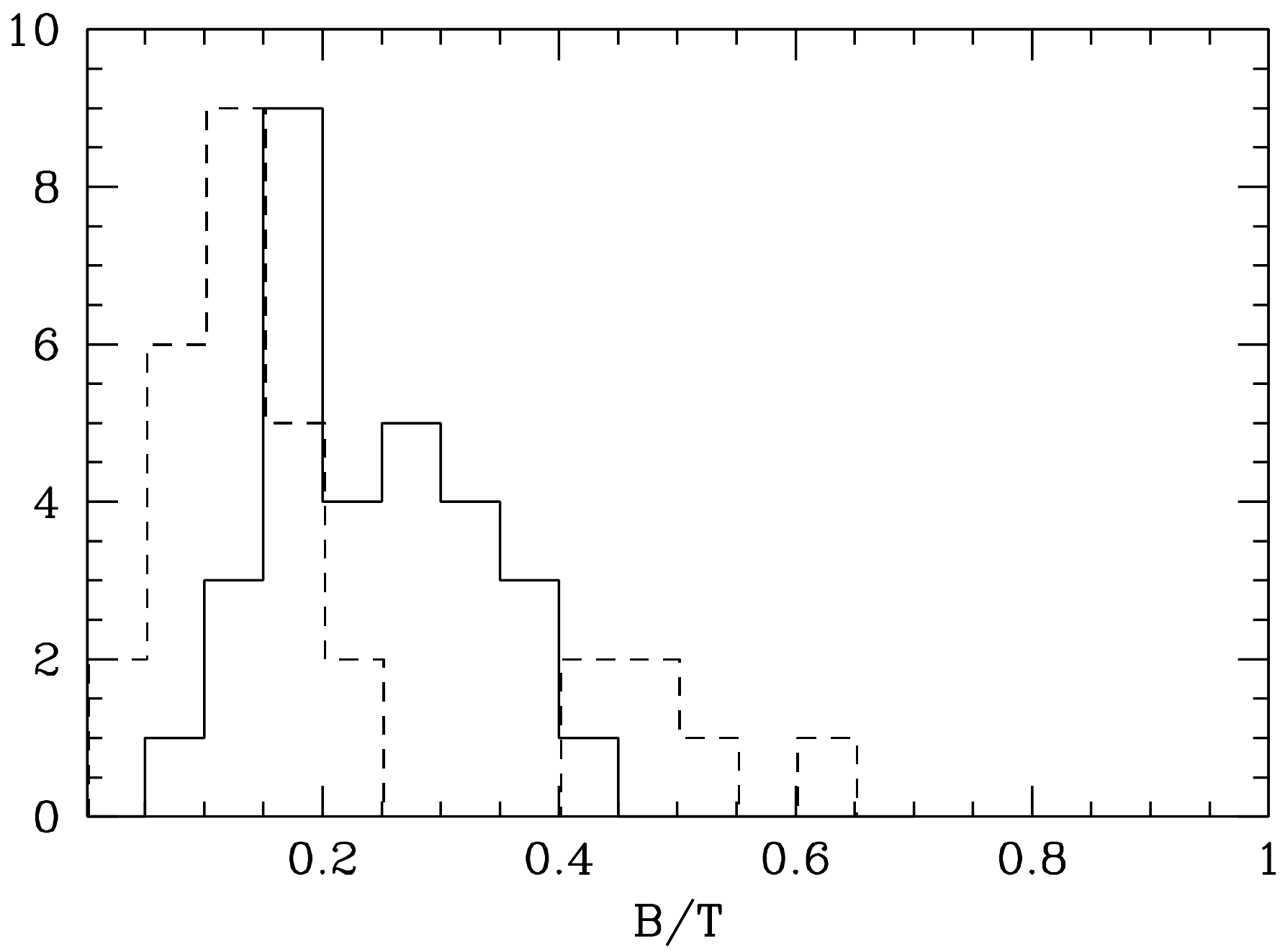}
\caption{Ratios of bulge to total stellar masses of Auriga galaxy simulations from the Gargiulo, et al. (2019) analysis. The distribution of their measure B/T$_{\rm sim}$  defined by the circularity parameter is plotted as the solid lines, and of B/T$_{\rm v}$ defined by the fit to equation~(\ref{eq:Sersic}) is plotted as the dashed lines.}\label{fig:B/Tmodel}
\end{center}
\end{figure}

\subsection{Distributions of Bulge to Total Ratios}\label{sec:BtoTratios-compared}

Fig.~\ref{fig:B/Tmodel} shows distributions of ratios of bulge to total stellar mass from the analysis by Gargiulo, Monachesi, G\'omez, et al. (2019) of the 30 model galaxies from the Auriga project (Grand, G\'omez, Marinacci, et al. 2017). These galaxies formed in dark haloes about as massive as the Milky Way, comparable to that of an \LS\ galaxy. The measure B/T$_{\rm sim}$ defines bulge star particles as those at galactocentric distances less than two effective bulge radii $R_{\rm eff}$ and with circularity parameters $\epsilon < 0.7$.  The ratio B/T$_{\rm v}$ is from the fit of the model galaxy surface brightness run normal to the disc to equation~(\ref{eq:Sersic}). The solid lines show the distribution of B/T$_{\rm sim}$ and the dashed lines the distribution of B/T$_{\rm v}$.  

It is encouraging that the observed and model distributions of bulge fractions B/T in Figs.~\ref{fig:B/Tdata} and~\ref{fig:B/Tmodel} are reasonably similar. It may be significant that the measure B/T$_{\rm v}$ based on the run of surface brightness with radius is closer to the observations, because the observations also make use of the fit to equation~(\ref{eq:Sersic}). And the S\'ersic indices of models and the nearby large spirals both tend to be characteristic of pseudobulges. 

There are two problems, however. First, the measure B/T$_{\rm sim}$ defines bulge star particles by the cut $\epsilon < 0.7$, meaning the orbits are far from circular (as illustrated in eq.~[\ref{eq:orbitexample}] and discussed further in Sec.~\ref{sec:epsilon_distributions}). This would be appropriate for classical bulges. But most of the local \LS\ spirals have pseudobulges, which tend to have small fractional departures of motions from the mean according to the arguments in Sec.~\ref{sec:bulges}.  Second, the bulge cut $\epsilon < 0.7$ leaves the considerable number of star particles with $\epsilon > 0.7$ with orbital eccentricities too large for the disk (as discussed in Sec.~\ref{sec:epsilon_distributions}) and too numerous for the stellar halo (Sec.~\ref{sec:HaloesandBulges}). By this line of argument  B/T$_{\rm sim}$ underestimates  bulge masses. But the distribution of  B/T$_{\rm sim}$  shown as the solid histogram in Figure~\ref{fig:B/Tmodel} already is seriously broader than the measurements in Figure~\ref{fig:B/Tdata}.

\begin{figure}
\begin{center}
\includegraphics[width=2.5in]{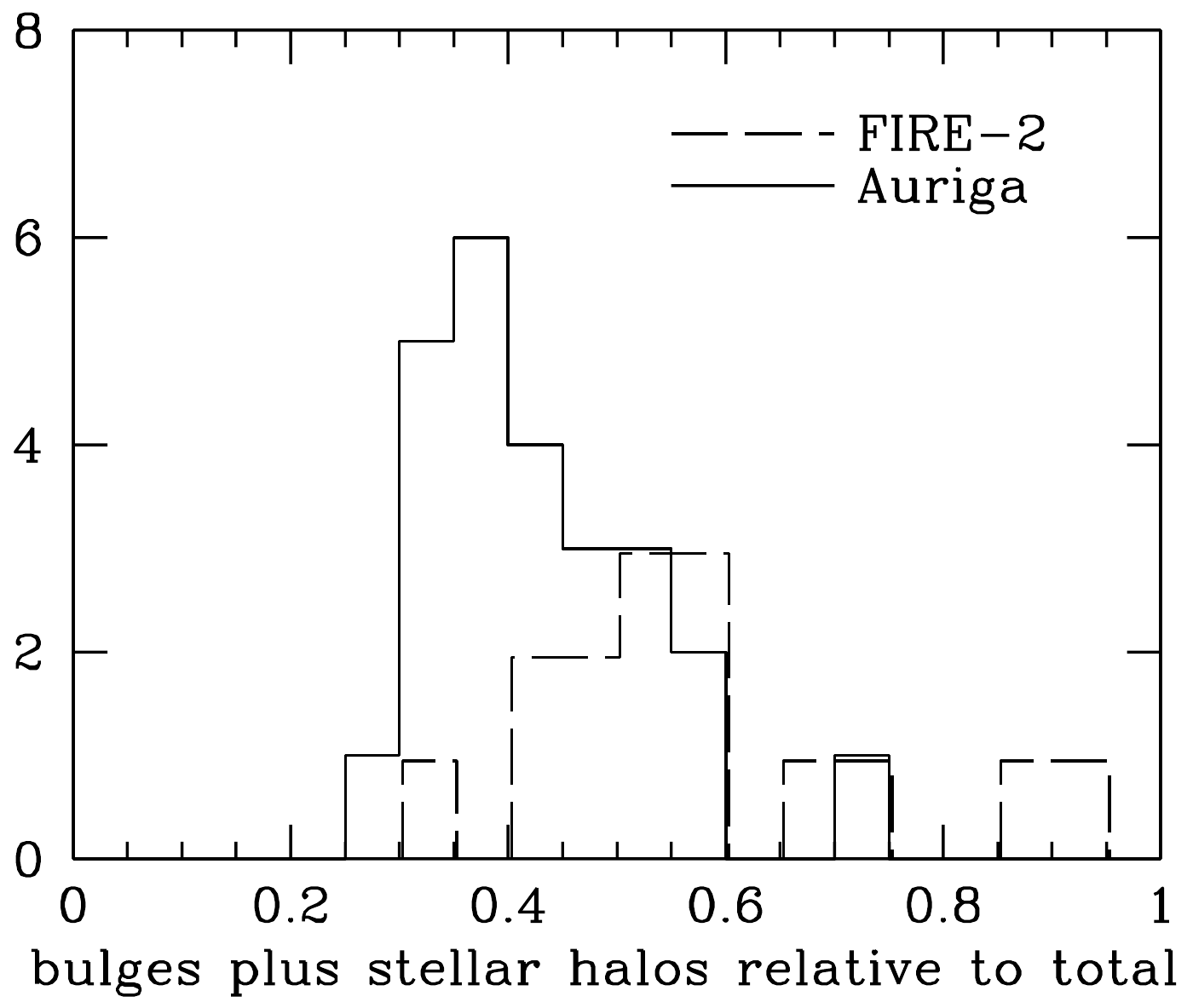}
\caption{Estimates of the bulge plus stellar halo mass fractions in Auriga and FIRE-2 galaxies.} \label{fig:BplusH}
\end{center}
\end{figure}

\subsection{Stellar Haloes Plus Bulges}\label{sec:HaloesandBulges}

Many nearby \LS\ spirals have quite small luminosity fractions in stellar haloes (Merritt, van Dokkum, Abraham, \& Zhang 2016; Harmsen, Monachesi, Bell, et al. 2017). Sanderson, Garrison-Kimmel, Wetzel, et al. (2018) point out the difficulty of applying these observations to test and constrain model galaxies. But a simple and I think reasonable measure starts from the mass of stars in orbits close enough to circular to qualify as parts of the disc or a pseudobulge. The remainder has to be the sum of the mass in a stellar halo and perhaps in a classical bulge. There is the problem that model bulges that are pseudobulges according to the S\'ersic index seem to have hot orbits according to the circularity index. So I take it that the star particles with circularity parameters too small to qualify for the disk are in the stellar halo or a bulge of some sort. 

Garrison-Kimmel, Hopkins, Wetzel et al. (2018) define the disc mass in a FIRE-2 model galaxy by the fraction of star particles with circularity parameters $\epsilon > 0.5$, along with a cut on galactocentric distance. The discussion in Section~\ref{sec:circularities} indicates that this allows a seriously hot distribution of star particles in the disc. So for the present discussion I take the disc star particles to have the more commonly used cut $\epsilon > 0.7$. The fraction in hot distributions of orbits in bulge plus stellar halo of some sort would then be 
\beq
(B+H)/T = 1-f^\ast_{\ge 0.7}.\label{eq:B+H/T}
\eeq
The fraction $f^\ast_{\ge 0.7}$ with $\epsilon \ge 0.7$ is listed in column 9 in table~1 in Garrison-Kimmel et al. (2018). The distribution of (B+H)/T defined this way for the 15 FIRE-2 galaxies is plotted in long dashes in Fig.~\ref{fig:BplusH}.

We can compare this distribution to the sum of estimates of bulge plus stellar halo masses in Auriga galaxies. I do not consider models Au29 and Au30, which Monachesi et al. do not find promising, and I exclude Au28, which does not look that much better. I add the bulge mass listed in the second column in table~1 in Gargiulo et al. (2019) to the {\it in situ} plus accreted halo masses in columns 5 and 6 in table~1 in Monachesi,  et al. (2019), and normalize the sum by the stellar masses listed in column 4 in Monachesi et al. The total stellar masses Gargiulo et al. 2019 use in their measure of B/T are in some cases a few tens of percent lower than what Monachesi et al. 2019 use, but the difference is small at the level of this discussion.  The definitions of stellar halo, bulge, and disk do not include all star particles. If these missing particles are not to be assigned to the disk then my estimate of (B+H)/T is a lower bound. 

The (B+H)/T distributions in the FIRE-2 and Auriga simulations are computed in different ways, but the approaches seem similar, and the two sets of models make reasonably similar predictions: the median value of the stellar mass fraction in bulges plus stellar haloes is about 
\beq
\hbox{(B+H)/T } = 0.45 \pm 0.05. \label{eq:BHoverT}
\eeq 

\begin{table}
\centering
\caption{Stellar Halo and Bulge Fractions in the Local Sample}
\label{tab:HBT}
\begin{tabular}{lcccc}
\hline
galaxy  & H/T  & B/T & (B+H)/T & type  \\
\hline
Milky Way & 0.01 & $0.19 \pm 0.02$ & 0.20 &P H \\
M31 & 0.15 & $ 0.32 \pm 0.02 $ & 0.47 &C H  \\
NGC4945 & 0.09 & $0.07 \pm 0.01$ & 0.16 &  P H \\
NGC2903 & $0.010\pm 0.007$ & 0.10 & 0.11 & P M\\ 
M81 & 0.02 & $0.34 \pm 0.02$ & 0.36 & C H\\
NGC253 & 0.08 & $0.15$ & 0.23 & P H\\
M101 & $0.001\pm  0.001$ & $0.03 \pm 0.01$ & 0.03 & P M\\
 M96 & $0.00\pm 0.03$ & $0.26$ & 0.26 &  P M  \\
M106 & $0.00\pm 0.02$ & $0.12 \pm 0.02$ & 0.12 & C  M   \\
NGC891 & 0.05 & -- &  -- &  -- H\\
 M95 & $0.00\pm 0.02$ & $0.16 $ & 0.16 & P M \\
\hline
\end{tabular}
\end{table}

Monachesi et al. (2019) point out that their Auriga stellar halo masses tend to be  considerably larger than the Harmsen et al. (2017) measurements for nearby galaxies. Another illustration of the observational situation is in Table~2, which collects measured halo fractions H/T of galaxies in the local \LS\ sample from Merritt et al. (2016) and Harmsen et al. (2017). The former used surface brightness measurements for the estimates in their table~1 of the fraction $f_{\rm halo}(>5R_h)$ of the stellar mass outside $5R_h$, where $R_h$ is the half-mass radius of the galaxy. Harmsen et al. used measured counts of detected red giant stars outside galactocentric distance 10~kpc with the assumption that the initial mass functions of stellar haloes are at least roughly similar to what is observed in our galaxy. Table~2 also includes the Harmsen et al. assessments of reasonable values for the Milky Way and M\,31. The three ellipticals in the local sample might be included in Table~2 with ${\rm (B+H)/T} = 1$. (I take the liberty of  aiding clarity by reducing the number of significant figures presented in the table.)

The third column in Table~2 is the bulge fractions B/T from Table~1. The fourth column is the sum of central values, and the last column lists the bulge type and the source of the halo measurement, M or H for Merritt et al. (2016) or Harmsen et al. (2017). 

The galaxy NGC\,7814 in the Harmsen et al. study is outside the local sample at  $D\sim 17$~Mpc, but worth noting because it resembles the Sombrero Galaxy, a mix of elliptical and spiral. The entry for the Sombrero Galaxy in Table~1 might be rewritten (B+H)/T $\sim 0.5$. Harmsen et al. find halo fraction H/T = 0.14 in NGC\,7814, and there seems to be room for an inner component at $r<10$~kpc for a total comparable to that of the Sombrero Galaxy. The galaxy NGC\,891 is in the local sample, and the Harmsen et al. (2017) halo fraction is entered in Table~2, but the bulge is obscured by dust and not measured. The central value of the measured distance to NGC\,4565 puts it just outside the local sample. Harmsen et al. find H/T = 0.03. Kormendy \& Bender (2019) argue that the bright central region of this galaxy is a pseudobulge, a bar seen close to edge on. 

The nearby face-on galaxy M\,101 shows little evidence of starlight in a bulge or  halo. Van Dokkum, Abraham, \& Merritt (2014) report that the ratio of the luminosity of the halo of this galaxy to its total luminosity is $H/T < 0.01$. Jang, de Jong, Holwerda, et al. (2020) put the fraction at H/T$\la 0.003$. Recall that the rotationally supported disc of this galaxy seems to run all the way in to a nuclear star cluster that is in effect a classical bulge with luminosity fraction B/T $\sim 10^{-4}$. This plus the faint stellar halo makes M\,101 a beautiful example of a most interesting phenomenon that I have not seen in models. 

The lesson from comparison of the measurements in Table~2 and the model predictions in Fig.~\ref{fig:BplusH} and equation~(\ref{eq:BHoverT}) is that many \LS\ galaxies have much smaller luminosity  fractions in stellar haloes plus bulges than might be expected from the models. The galaxies M\,31 and M\,81 have large bulge plus stellar halo fractions, (B+H)/T = 0.46 and 0.36. All the rest in this still limited sample have (B+H)/T $\la 0.25$. The Auriga and FIRE-2 simulations have no examples at (B+H)/T $< 0.25$. The theory seems to be well separated from the observations. This is one of the five challenges discussed in Section~\ref{sec:challenges}.

We might pause to consider whether the stellar halo and classical bulge of a spiral galaxy are parts of the same phenomenon, artificially separated by an observationally convenient cut in surface brightness or distance from the center of the galaxy. The usual thinking is that the large bulge and stellar halo of M\,31 are the results of quite significant mergers. But M\,81 has a substantial classical bulge and a much more modest stellar halo. If pseudobulges have cold distributions of orbits, and stellar haloes hot, then one might not expect to find a correlation of halo and pseudobulge luminosities. Indeed, images of the galaxy NGC\,253 on the web look wonderfully flat, but Harmsen et al. (2017) assign it a considerably more luminous halo than M\,101. It may be significant, however, that Table~2 includes examples of low luminosity fractions in both pseudobulge and stellar halo. 

\subsection{Circularity Parameters and Velocity Dispersions}\label{sec:epsilon_distributions}

The star particle circularity parameter diagnostic $\epsilon$ is discussed in Section~\ref{sec:circularities}. If the  bulge and stellar halo in a simulation are not rotating then the mass in these components may be taken to be double the count of star particles with negative $\epsilon$. For example, by this measure the two models presented by Murante, Monaco, Borgani, et al. (2015) have B/T = 0.20 and 0.23. But the authors are not claiming that these are useful measures of bulge masses. Their distributions of $\epsilon$ have the same broad band as in the Guedes, et al.  (2013)  Eris model in the interval $0\la\epsilon\la 0.7$. It indicates a considerable mass fraction in something like a bulge with substantially less than full rotational support. 

There has been impressive progress since Eris in the spatial and mass resolutions of simulations of galaxy formation, and in the modeling of the complexities of the evolution, by a variety of groups. And it is notable that the distributions of $\epsilon$ continue to show the large mass fractions that do not seem to belong in a realistic disc or pseudobulge. In the recent example from Kretschmer, Agertz, \& Teyssier (2020) the distribution of $\epsilon$ in the left-hand panel in their fig.~5 looks quite like the Eris distribution from seven years earlier. The less familiar-looking  distribution in the right-hand panel of Kretschmer et al. might be a useful approximation to an irregular galaxy, or maybe a first step to an elliptical. 

Most of the distributions of $\epsilon$ in the 30 Auriga model galaxies (in fig.~7 in Grand et al. 2017) have a local peak or discontinuous change of slope at $\epsilon =0$ and a prominent peak at $\epsilon = 1$. These features are expected in a realistic spiral galaxy with a classical bulge. However, the models typically include a considerable mass fraction with circularity parameters in the range $0.1\la\epsilon\la 0.7$. This is indicative of substantial departures from the circular motions expected in the disk and maybe the pseudobulge of an \LS\ galaxy.  

The distributions of $\epsilon$ in the Buck, Obreja, Macci\`o  et al. (2019) NIHAO Ultra High Definition suite (their fig. 10) have the usual significant mass fractions at $0.1\la\epsilon\la 0.7$. The same is true of the mean in the EAGLE simulations (fig. 14 in Trayford, Frenk, Theuns, et al. 2019) at low redshift. 

Among the FIRE-2 galaxy simulations  by Garrison-Kimmel, et al. (2018) some of the distributions of the circularity parameter in their fig.~1 have at most a slight feature at $\epsilon=0$. The authors term these galaxies ``nearly bulgeless.'' In the two most pronounced examples, the models named Romeo and Juliet, the authors assign disc fractions $f_{\rm disc}^*=0.8$ based on a disc cut at $\epsilon >0.5$. But  in these two galaxies 35 and 41 per cent of the model star particles are at $\epsilon<0.7$. This is a considerable departure from circular orbits, as illustrated in Fig.~\ref{fig:orbit}. There does not seem to be room for this much mass in stars moving with large departures from rotational support in the observed nearly bulgeless galaxies in the local sample. 

I offer for comparison two examples of star velocity dispersions in spiral galaxies. The Kregel, van der Kruit \& Freeman (2004) measurements of the radial rms velocity dispersion $\sigma_r$ in the disks of spiral galaxies indicate that the ratio of the radial dispersion to the circular velocity $v_c$ is on average
\beq
\sigma_r/v_c = 0.3 \hbox{ at } r=h,\quad \sigma_r/v_c = 0.2 \hbox{ at } r=2h.
\label{eq:measured-dispersions}
\eeq
The Kregel et al. measurement is the dispersion at one radial scale length $h$; the extrapolation to two scale lengths, using $\sigma_r\propto e^{-r/2h}$, allows comparison to the  dispersion at about our position in the Milky Way. The local radial rms star velocity distribution weighted by star counts in the thin and thick disks is (from Anguiano, Majeweski, Freeman, et al. 2018, 2020, who find thin and thick radial dispersions $\sigma_r=33$ and $57\hbox{ km s}^{-1}$ with star count fractions 0.82 and 0.17, and their adopted circular velocity $v_c=240$ km s$^{-1}$)
\beq
\sigma_r/v_c=0.16.\label{eq:local-dispersion}
\eeq

Equations (\ref{eq:measured-dispersions}) and (\ref{eq:local-dispersion}) can be compared to the ratio $\sigma_r/v_c(\epsilon)$ as a function of the circularity parameter $\epsilon$ in the flat rotation curve model in Section~\ref{sec:circularities}: 
\beq
{\sigma_r\over v_c}(\epsilon = 0.9) = 0.31,\quad 
{\sigma_r\over v_c}(0.8) = 0.43,\quad {\sigma_r\over v_c}(0.7) = 0.53.
\label{eq:model-dispersion}
\eeq
Theory and observation are reconciled only if the circularity parameter of disk stars is taken to be considerably closer to unity than a usual choice, $\epsilon > 0.7$. Another indicator of the problem is the central velocity dispersions of Auriga disk stars normal to the disk, $\sigma_z\sim 0.4v_c$ (fig.~10 in Grand et al. 2017). The circularity parameter places only a broad constraint on motion normal to the disk, but the similarity of velocity dispersions normal to the disk and in the radial direction (eq.~\ref{eq:orbitexample}) checks the indication of uncomfortably hot distributions of star particle orbits in model galaxies.

The solid lines in Figure~\ref{fig:B/Tmodel} show the distribution of Auriga galaxy bulge to total star mass ratios B/T when the bulge stars are defined by the cut $\epsilon < 0.7$ (with a reasonable cut on galactocentric radius). But that leaves for the disk the star particles with $\epsilon > 0.7$ that have orbital eccentricities much larger the measurements in equations~(\ref{eq:measured-dispersions}) and~(\ref{eq:local-dispersion}). Reassigning the star particles with large radial velocity dispersions from the disk to a classical bulge would increase the difference between the observations  of B/T plotted in in Figure~\ref{fig:B/Tdata} and the model values shown as the solid histogram in Figure~\ref{fig:B/Tmodel}. The model distribution plotted as the dashed lines in Figure~\ref{fig:B/Tmodel} is closer to the observations, but that would have still larger velocity dispersions in the disks. 

\section{The Bimodal Field Galaxy Population}\label{sec:bimodality}
 
The two great morphology classes of \LS\ galaxies are spirals and ellipticals. Figure~\ref{fig:B/Tdata} is the illustration in the local sample. Bluck, Bottrell, Teimoorinia, et al. (2019) show a broader illustration in their fig.2.  Among the most luminous galaxies at optical wavelengths, $L\sim 10 L_\ast$, most are ellipticals or cDs, but detections in the Sloan Digital Sky Survey reveal that some 6\% are giant spirals (Ogle, Lanz, Appleton, et al. 2019). And there are many nearby examples of the two morphologies at $L\sim 0.1 L_\ast$. This early-late bimodality is observed in colors, element abundances, interstellar matter, and stellar formation rates and evolution ages. But the bimodality of morphologies is not a subtle effect, and surely an important guide to how the galaxies formed. 

There are complicating details. The Sombrero Galaxy M\,104 is a striking example of a mixed spiral and elliptical, and another, NGC\,7814, is not far outside the local sample. There are the varieties of S0 galaxies. NGC\,3115 is a large relatively nearby example; the Local Universe catalog puts it at $D\sim  11$~Mpc. The S0 galaxy NGC\,5102 is even closer, at $D\sim  4$~Mpc. Its luminosity is about half the luminosity cut for Table~\ref{tab:nearbysample}. The S0 galaxies in clusters may be normal spirals that were stripped of gas by the ram pressure of the intracluster plasma. But since NGC\,3115 and NGC\,5102 are not near a cluster of galaxies how were they so neatly stripped of gas? Though the first-ranked  members of clusters tend to be cDs, Li \& Chen (2019) find that in some less rich clusters they are unclassifiable, or resemble spirals. And some galaxies are exceedingly  luminous in the infrared. All of these kinds of objects have something to teach us about how the galaxies formed, but all are less common than \LS\ galaxies, and not part of the considerations of what we might learn from the local sample. There is an exception, however: merging spirals (Schweizer 1990). Lah{\'e}n, Johansson, Rantala, et al. (2018) argue that in about 3~Gyr the Antennae Galaxies will resemble the elliptical M\,105 in Table~1. Naab \& Ostriker (2009) dispute this; they see problems with the patterns of chemical abundances. But mergers produce galaxies of some sort. A far future version of Table~1 might include such galaxies produced by mergers of the blobs associated with the galaxies M\,51a and NGC\,4490.

What determines whether a protogalaxy precursor to an \LS\ galaxy in the field outside rich clusters will become a spiral or an elliptical? The proposition that ellipticals grow by dry mergers only changes the question to why the mergers are dry in some haloes in the field, wet in others. Since the global mean ratio of ellipticals to spirals increases with increasing stellar mass (Tasca \& White  2011; Bluck, et al. 2019, fig.2), the protogalaxy mass is an important determining factor (Johansson, Naab, \& Ostriker 2012; Clauwens, Schaye, Franx, \& Bower 2018). Consistent with this, the three ellipticals in the local sample have luminosities $L_K = 0.7\times 10^{11}$ to $2\times 10^{11}$, toward the upper end of the range of values of $L_K$ in the local sample. But in the local sample the three spirals NGC 253, 6744, and 6946, which look like rotationally supported discs with modest bulges, have luminosities in the same range as the three ellipticals. And recall that there are ellipticals as well as spirals at $L\sim 0.1 L_\ast$ to $L\sim 10 L_\ast$. This is a true bimodality. 

The Fall \& Romanowsky (2018) relation among galaxy stellar mass, angular momentum, and bulge fraction shows the importance of angular momentum. The tidal torque picture for the origin of the rotation of galaxies does not appear to be  likely to have produced a bimodality of galaxy types, however. 

This evidence indicates a bistability somewhere in the course of evolution of cosmic structure. The distributions of (B+H)/T in the model galaxies in Fig.~\ref{fig:BplusH} suggest a single peak with tails toward the two most common types, near pure disc spirals and disc-free ellipticals. These simulations do not seem to have captured the bistability, whatever it is. Section~\ref{sec:ELS} offers thoughts on what might have happened.

\section{Five Challenges}\label{sec:challenges}

I offer a summary of these considerations in the form of challenges that seem reasonably well founded and possibly simple enough for productive contemplation. Three cautions are to be noted. This discussion compares stellar mass fractions in bulges in model galaxies to luminosity fractions in observed galaxies. I cannot assess how much room for error that allows. The roughly 30 galaxies (depending on how you count) in the local sample in Table~1 have been characterized by seasoned observers, but a second point to bear in mind is my assumption that these galaxies constitute a fair sample of the \LS\ population outside clusters. Third, I am assuming that pseudobulge stars move with velocity dispersion smaller than the mean. I do not know whether that may only be true for some subset of pseudobulges.

1. Pseudobulges. Model galaxies are said to tend to have pseudobulges because fits of the exponential plus S\'ersic function in equation~(\ref{eq:Sersic}) tend to indicate S\'ersic index $n<2$, closer to the run of surface brightness in a disk than an elliptical. But in the model pseudobulges discussed here the distributions of star particle circularity parameters indicate the considerable departures from circular motions illustrated in equation~(\ref{eq:orbitexample}). When a bar is present the star particles may have small values of $\epsilon$ because they are moving in organized noncircular patterns. Otherwise it seems that the motions of stars in model pseudobulges are hot. That not what would be expected of pseudobulges from the considerations in Sec.~\ref{BulgesPseudobulges}. And we see in Table~1 that pseudobulges are common among nearby \LS\ galaxies. The challenge is either to explain why simulations fail to predict the cold flows of stars in the pseudobulges common in \LS\ galaxies, or else to explain why the notion of well-ordered flows of pseudobulge stars is wrong. 

2. Bulge Fractions. If the run of model surface brightness fitted to the exponential plus S\'ersic function yields a good measure of the bulge fraction then the distribution of Auriga bulge fractions B/T$_{\rm v}$ in Fig.~\ref{fig:B/Tmodel} is not far from the observed distribution in Fig.~\ref{fig:B/Tdata}, within reasonable uncertainties in theory and observation. But there is the serious problem discussed in Section~\ref{sec:BtoTratios-compared}. The distribution B/T$_{\rm sim}$ based on the cut $\epsilon< 0.7$ is significantly further from the observations. And B/T$_{\rm sim}$ seems to be an underestimate because it assigns a distribution of hot orbits to the disk. The more natural home for star particles with $0.7\la\epsilon\la 0.9$ is a classical bulge or the inner parts of a stellar halo. How might the model bulge fractions be reduced to better fit the evidence? Guedes et al. (2013) fig.~3 and Gargiulo et al. (2019) fig. 5 show models in which most bulge stars formed at about the same range of galactocentric distances that they are now. Governato et al. (2010) point out that if B/T is too large for these stars it may be because too much mass in diffuse baryons made its way to the central regions before conversion to stars. The generations of models since then may not have been specifically designed to reduce this mass transfer, and the typical values of B/T, but the goal is implicit in the program of improving model galaxies. And the challenge of reconciling model bulge mass fractions with observed bulge luminosity fractions seems to remain open. 

3. Stellar Haloes. The evidence reviewed in Sec.~\ref{sec:HaloesandBulges} is that the stellar halo mass fractions H/T in models tend to be considerably larger than the luminosity fractions H/T in nearby \LS\ spiral  galaxies. The observed sample is not large and may not be representative, but the discrepancy merits serious consideration. The substantial mass fraction in a model stellar halo is at least in part a result of the prediction of the standard $\Lambda$CDM cosmology that galaxies grew by merging of a hierarchy of subhaloes within subhaloes. Stars that formed in subhaloes before merging with the main halo of the growing galaxy would seldom end up joining a cold flow in the disc or in the adopted picture of a pseudobulge; they are far more likely to contribute to the stellar halo. Substantial fractions of stars in the  haloes of Auriga galaxies formed outside identified gravitationally bound subhaloes. But these {\it in situ} stars had to have formed out of gravitationally bound clouds: perhaps subhaloes too small to be identified; perhaps diffuse baryons tidally stripped from subhaloes. If a protogalaxy grew by the merging of matter with this significant substructure then the challenge would be to show how the baryon concentrations large and small could have ``known'' which was to be the single one in which nearly all the visible stars would be forming, so as to place  an acceptably small fraction of star particles in the stellar halo. 

4. Velocity Dispersions in Disks. There is a trade-off between the masses of model spiral galaxy bulges and the velocity dispersions in their  disks. In the simple model in equation~(\ref{eq:orbiteq}) the circularity parameter $\epsilon = 0.9$ translates to rms radial velocity dispersion $\sim 30\%$ of the circular velocity (eq.~[\ref{eq:orbitexample2}]). This seems to be marginally acceptable according to the measures in equations~(\ref{eq:measured-dispersions}) and~(\ref{eq:local-dispersion}). But model disk star particles commonly are taken to be those with $\epsilon > 0.7$ and the still larger departures from circular orbits illustrated in Figure~\ref{fig:orbit} and equation~(\ref{eq:model-dispersion}). Perhaps departures from axial symmetry and conservation of orbit energy inflate the apparent departures from quiet flow derived from equation~(\ref{eq:orbiteq}). That could be checked by statistics of velocity dispersions relative to circular velocities of star particles in model disks. Or perhaps the approximations required to deal with the complexities of star formation introduce artificially large dispersions of star velocities at formation. But the improvements of simulations since Eris (Guedes, et al. 2013) have failed to change the apparent prediction of large velocity dispersions in model disks. Thus the challenge remans to demonstrate consistency of star particle velocity dispersions in model disks with what is known about velocity dispersions of stars in the disks of \LS\ spiral galaxies.

5. The Spiral--Elliptical Bimodality. In the field outside rich clusters, at optical luminosities $L\sim 0.1 L_\ast$ to $L\sim 10 L_\ast$, the two great classes of galaxy morphology are spirals with modest stellar haloes and ellipticals with modest disks. There are irregular galaxies of many sorts, but they are less common. The challenge and opportunity is to identify the bistability that caused this somewhere along the course of evolution of cosmic structure.   

\subsection{What do the Challenges Suggest?}\label{sec:ELS}

The phenomenology suggests that many \LS\ galaxies grew by a gentle rain of diffuse matter that dissipatively settled into rotational support in a single growing disc before being incorporated in stars. This brings to mind the Eggen, Lynden-Bell \& Sandage (1962) picture for formation of the Milky Way by a roughly monolithic collapse of initially diffuse baryons. Eggen et al. pointed out that the contracting mass distribution would have to have had enough substructure to have allowed some star formation as matter was settling, so as to account for the high-velocity stars. But weaker subclustering than the $\Lambda$CDM prediction would help avoid accumulation of an unacceptably luminous stellar halo. And a scarcity of stars in the diffuse matter as it was settling would help avoid disturbing the growing disc (e.g. T\'oth \& Ostriker 1992; Kazantzidis, Zentner, Kravtsov, et al. 2009). 

Galaxies do merge, and mergers can change morphologies (Toomre 1977). While M\,101 seems to have suffered no significant accretion of subhaloes containing stars, it is reasonable to add to the phenomenological picture the standard idea that M\,31 and M\,81 owe their classical bulges to more serious mergers after formation of their first generations of stars.  The idea that ellipticals were assembled by dry mergers is supported by the shells seen in some ellipticals, and suggested by the concentration of early-type dwarfs in the outskirts of the elliptical galaxy Centaurus A (Karachentsev, Sharina, Dolphin, et al. 2002; Crnojevi\'c, Grebel, \& Koch 2010). But the phenomenology allows us to imagine instead that the elliptical in the Centaurus group formed by an early merger of two nearly monolithic haloes that happened to be unusually close to each other. This merger would have to have been violent enough to have triggered rapid star formation, which could have scattered debris, producing the early-type satellites of Centaurus A. The other large galaxy in this group, M\,83, with its pseudobulge and the usual concentration of mostly late-types satellites, would have escaped violent mergers, instead growing out of wet mergers and accretion. The presence of satellites requires departures from monolithic protogalaxies, as does the rather substantial stellar halo around the near pure disc galaxy NGC\,253, but perhaps lesser departures than in the conventional $\Lambda$CDM theory. 

A consideration reviewed in Peebles \& Nusser (2010), and to be added to the issues, is the evidence that the properties of \LS\ galaxies are insensitive to environment. This is a starting assumption for the Halo Occupation Distribution model (Berlind, Weinberg, Benson, et al. 2003), but it is curious. Apart from occasional mergers, \LS\ galaxies seem to have evolved as island universes, independent of the environment, yet the ratio of early to late types of galaxies is a function of environment. Perhaps that is because violent mergers at high redshift that produced early-type galaxies were more frequent in higher density regions. What would we make of less violent mergers? That is among the considerations that would have to be addressed by simulations of a model that encourages more nearly monolithic galaxy formation.

\section{Adjusting Initial Conditions}\label{AdjustingCosmology}

The $\Lambda$CDM cosmology certainly might be improved by a more realistic model for the dark sector. But the simpler idea considered here is an adjustment of initial conditions. 

In the warm dark matter (WDM) model, primeval mass density fluctuations are suppressed on scales less than a chosen comoving value $M_{\rm WDM}$. The idea has a long history (Blumenthal, Pagels, \& Primack 1982; Bond, Szalay, \& Turner 1982), it still is discussed (e.g. Adhikari, Agostini, Ky, et al. 2017; Lovell, Hellwing, Ludlow, et al.\ 2020; Leo, Theuns, Baugh, et al. 2020), and it can suppress substructure within protogalaxies. It still allows unwanted promiscuous merging on scales $\sim M_{\rm WDM}$, however. I have experimented with remedying this by changing the shape of the primeval mass fluctuation power spectrum on scales $\sim M_{\rm WDM}$. But the even simpler adjustment presented here keeps an approximation to the WDM power spectrum while making the primeval mass density fluctuations non-Gaussian.

There are well-discussed challenges to WDM from observations of the Lyman-$\alpha$ forest and the shapes of the dark matter distributions around galaxies. The former might be fixed by a mixed dark matter model, another old idea (Davis, Lecar, Pryor, \& Witten 1981) that still is discussed (e.g. Adhikari, Agostini, Ky, et al. 2017). Assessment of the latter depends on the effect of the non-Gaussianity to be discussed, which would require numerical simulations. The non-Gaussian initial conditions could be challenged by the near Gaussian nature of the cosmic microwave background anisotropy, except that the relevant length scales are very different. Maybe non-Gaussianity is a function of scale. 

\subsection{Non-Gaussian Initial Conditions}\label{sec:non-G}

Consider the primeval mass density contrast $\delta(x)$ as a function of position $x$ along a straight line through a realization of a scale-invariant process with the WDM cutoff. Let $\delta_{\rm G}(x)$ be the usual WDM Gaussian realization, and consider two adjustments of initial conditions:
\beq
\delta(x) = {{\delta_{\rm G}(x) + F\left(\delta_{\rm G}(x)^2\langle\delta_{\rm G}^2\rangle^{-1/2} - \langle\delta_{\rm G}^2\rangle^{1/2}\right)}\over{(1 + 2F^2)^{1/2}}},\label{eq:nong-2}
\eeq
\beq
\delta(x) = {{\delta_{\rm G}(x) + F\delta_{\rm G}(x)^3/\langle\delta_{\rm G}^2\rangle}\over {(1 + 6F + 15F^2)^{1/2}}}.\label{eq:nong-3}
\eeq
The constant $F$ is to be chosen for the wanted degree of skewness or excess kurtosis. The processes $\delta$ and $\delta_{\rm G}$ have zero means and are normalized to the same variance. The quadratic expression is adapted from a test for non-Gaussianity of the cosmic microwave background radiation (Verde, Wang, Heavens, \& Kamionkowski 2000; Komatsu \& Spergel 2001). The cube of $\delta_{\rm G}(x)$ in the second expression is a convenient alternative. 

The two-point correlation function offers a simple measure of the effect of the added terms on the mass fluctuation spectrum. The correlation function in the quadratic model in equation~(\ref{eq:nong-2}) is
\beq
\langle\delta_1\delta_2\rangle = 
\langle\delta_1\delta_2\rangle_{\rm G} \left[
{1+2F^2\langle\delta_1\delta_2\rangle_{\rm G}/\langle\delta^2\rangle_{\rm G}}\over {1+2F^2}\right],
\eeq
where $\langle\delta_1\delta_2\rangle_{\rm G}$ is the two-point function for the Gaussian process. The two-point function for the cubic  model in equation~(\ref{eq:nong-3}) is 
\beq
\langle\delta_1\delta_2\rangle =  \langle\delta_1\delta_2\rangle_{\rm G} \left[ 
{1+6F + 9F^2+ 6F^2\langle\delta_1\delta_2\rangle_{\rm G}^2/\langle\delta^2\rangle_{\rm G}^2}\over {1+6F+15F^2}\right].
\eeq
At $F\ll 1$ the functions are changed from that of $\delta_{\rm G}$ only by terms of order $F^2$. 

I hesitate to apply equations~(\ref{eq:nong-2}) or~(\ref{eq:nong-3}) to a near scale-invariant power law power spectrum truncated at some very small scale because that places considerable substructure within protogalaxies, which may encourage excess star formation prior to merging, and excess mass in a hot distribution of orbits. But it certainly might be considered.

\begin{figure}
\begin{center}
\includegraphics[angle=0,width=3.5in]{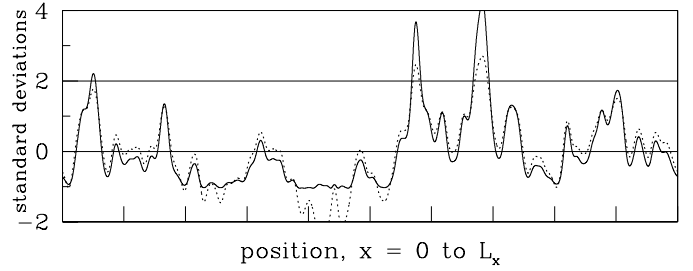} 
\caption{The dotted curve is a realization of the Gaussian process $\delta_{\rm G}$. The solid curve is the primeval density contrast $\delta(x)$ with the quadratic term in eq.~(\ref{eq:nong-2}) and $F=0.3$. The vertical axis is the contrast in units of standard deviations of $\delta_{\rm G}$.}\label{fig:plot_quadratic}
\end{center}
\end{figure}

\begin{figure}
\begin{center}
\includegraphics[angle=0,width=3.5in]{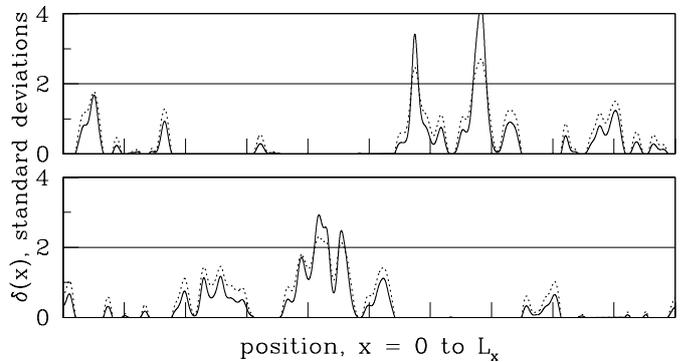}
\caption{As in Fig.~\ref{fig:plot_quadratic} for the cubic term in eq.~(\ref{eq:nong-3}) and the same $\delta_{\rm G}(x)$ and $F$.}\label{fig:plot_cubic}
\end{center}
\end{figure}

In the examples in Figs.~\ref{fig:plot_quadratic} and~\ref{fig:plot_cubic} the Gaussian function with period $L_x$ along the line $x$ is computed as  
\beq
\delta_{\rm G}(x) = 2\sum_{k\geq 1}^{k \gg k_c} \delta_k \cos (2\pi x k/L_x + \phi_k), \
\delta_k \propto \exp^{-k^2/2k_c^2}/\sqrt{k}.
\eeq
The normalization of $\delta_k$ does not affect the ratios of the Gaussian and non-Gaussian terms in equations~(\ref{eq:nong-2} and~(\ref{eq:nong-3}). The phases $\phi_k$ are random, as usual. The length of the plot is $L_x$ in units we may choose. (In the computation $L_x=10^4$ and $k$ is the positive integers.) The exponential factor in $\delta_k$ approximates the WDM cutoff of the power spectrum at the length scale $\sim L_x/k_c$. In the two examples $k_c=30$, meaning there are roughly 30 oscillations across the widths of the figures. The factor $k^{-1/2}$ in $\delta_k$ makes the fluctuations scale-invariant along the line on scales $\gg L_x/k_c$.

The dotted curves in Figs.~\ref{fig:plot_quadratic} and~\ref{fig:plot_cubic}  are the same realization of the random Gaussian process, and the solid curves are the non-Gaussian processes in equations~(\ref{eq:nong-2}) and~(\ref{eq:nong-3}) with the same non-Gaussian parameter $F=0.3$. Since realizations of the cubic model are statistically unchanged by a change of sign of $\delta$ we get two examples by plotting only positive values in one panel of Fig.~\ref{fig:plot_cubic} and only negative values, with the sign changed, in the other. 

The skewness of initial conditions in Fig.~\ref{fig:plot_quadratic} and the excess kurtosis in Fig.~\ref{fig:plot_cubic} both have the effect of increasing 2$\sigma$ upward density fluctuations. This is in the direction of the Eggen, Lynden-Bell, \& Sandage (1962) picture, which I have argued is suggested by the observations. The cubic model in Fig.~\ref{fig:plot_cubic} is closer to what seems to be indicated, because it increases density fluctuations above the 2$\sigma$ line and decreases them below the line. That is, it suppresses subclustering around a growing mass concentration, which may offer a better approximation to a real protogalaxy. It may even help account for the presence of hosts for quasars at high redshift, as in downsizing (Cowie, Songaila, Hu, \& Cohen 1996), and for the impressive scarcity of dwarf galaxies in the Local Void (Peebles \& Nusser 2010 fig. 1).

\section{Concluding Remarks}

We must be cautious about adjusting a theory to fit what is wanted, which may be meaningless. But recall that CDM and then Einstein's cosmological constant were added to the cosmological model to make the theory fit reasonably persuasive evidence. I have argued that there is reasonably persuasive evidence that simulations of galaxy formation based on the  $\Lambda$CDM theory with Gaussian initial conditions produce unacceptably large fractions of stars in hot distributions of orbits. It is appropriate to seek an adjustment of the theory that might relieve the problem, and natural to look first at the sub-grid physics. But this has been examined in several generations of models by several groups. The stability of the gross form of the distribution of the circularity parameter $\epsilon$, with the substantial mass fraction in what looks like hot distributions of orbits, suggests this is characteristic of $\Lambda$CDM  in a considerable range of ways to treat the baryon physics. The adjustment of initial conditions proposed here is more contrived than the introductions of  CDM and $\Lambda$, but it is in the same spirit. Whether it would remedy any of the challenges listed in Section~\ref{sec:challenges} remains to be examined,  perhaps by numerical simulations already run with power law Gaussian initial conditions and run again with a degree of departure from power law and Gaussianity. 

\section*{Acknowledgements}

I am grateful for informative discussions with  Dimitri Gadotti, Fabio Governato, Rob Grand, Patrick Ogle, Piero Madau, Michael Strauss, and Simon White.

\bsp	
\label{lastpage}
\end{document}